\documentclass{article} \usepackage{slashed,feynmf,epsfig,amsmath,amssymb,enumitem} \usepackage[papersize={8.5in,11in}]{geometry}
  \renewcommand{\vec}[1]{\boldsymbol{\mathrm{#1}}} \usepackage{bm}
\begin{document} 
\title{LIGO GW150914 and GW151226 Gravitational Wave Detection and Generalized Gravitation Theory (MOG)} 
\author{J. W. Moffat\\~\\
Perimeter Institute for Theoretical Physics, Waterloo, Ontario N2L 2Y5, Canada\\
and\\
Department of Physics and Astronomy, University of Waterloo, Waterloo,\\
Ontario N2L 3G1, Canada}
\maketitle



\begin{abstract}
The nature of gravitational waves in a generalized gravitation theory is investigated. The linearized field equations and the metric tensor quadrupole moment power and the decrease in radius of an inspiralling binary system of two compact objects are derived. The generalized Kerr metric describing a spinning black hole is determined by its mass $M$ and the spin parameter $a=cS/GM^2$.  The LIGO-Virgo collaboration data is fitted with smaller binary black hole masses in agreement with the current electromagnetic, observed X-ray binary upper bound for a black hole mass, $M\lesssim 10M_\odot$. 
\end{abstract}

\maketitle


\section{Introduction}

A century after the fundamental prediction by Einstein, based on his gravitational field equations~\cite{Einstein1} of gravitational waves~\cite{Einstein2,Einstein3}, and Schwarzschild's derivation of his solution to the general relativity (GR) field equations~\cite{Schwarzschild}, which led to the prediction of black holes, the first direct detection of gravitational waves has been reported~\cite{Abbott1,Abbott2}. The gravitational waves are produced by the merging of a binary black hole system to form a single black hole. The measurements lead to a new access to the physical properties of spacetime and strong field gravity. The observations of the energy loss by Taylor and Weisberg~\cite{TaylorWeisberg}, following the discovery of the binary pulsar system PSR B1913+16 by Taylor and Hulse~\cite{HulseTaylor} demonstrated the existence of gravitational waves. 

In the following, we will investigate the nature of gravitational waves in a generalized gravitation theory called Scalar-Tensor-Vector-Gravity (STVG), also known in the literature as MOG (modified gravity)~\cite{Moffat1}. The theory has been studied as an alternative to GR without (detectable) dark matter in the present universe, and fits to galaxy rotation curves and galaxy clusters have been obtained~\cite{MoffatRahvar1,MoffatRahvar2,MoffatToth}. Moreover, the theory has been applied to cosmology with an explanation for the growth of structure in the early universe and fits have been obtained to the CMB data~\cite{Moffat3,Moffat4}. In the early universe cosmology, the mass of the vector field $\phi_\mu$ is $m_\phi\gtrsim 10^{-22}$ eV and acts as a cold dark matter particle with density $\rho_\phi >\rho_b$, where $\rho_\phi$ and $\rho_b$ denote the density of the boson particle and baryons, respectively. As the universe expands and enters the era of the formation of stars and galaxies the effective boson mass decreases to the value $m_\phi\sim 10^{-28}$ eV, and due to the weak gravitational coupling of the particle to ordinary matter the particle will be unobservable in the late-time universe\footnote[1]. 

An alternative early universe cosmology without dark matter is formulated in~\cite{TothMoffat}.

In our generalized gravitational theory electromagnetic waves (photons) and gravitational waves (gravitons) move with the speed of light.
The null geodesic equation for photon paths is determined in a Jordan frame conformal metric, and in the Einstein frame metric the gravitational constant for photon paths is screened, yielding the deflection of light by the Sun, and the Shapiro time delay in agreement with GR. The enhanced gravitational interaction experienced by photons in the lensing of galaxies and galactic clusters leads to an explanation of gravitational lensing data without dark matter~\cite{Moffat2}. 

The paper is organized, as follows. In Section 2, we present the STVG field equations, while in Section 3, we investigate the linearized weak field approximation of the field equations, the modified two-body acceleration law and the tensor gravitational wave equations for a binary system.  Section 4, presents the generalized Kerr solution of the gravielectric vacuum STVG field equations, while in Section 5, we investigate the inspiralling merger of two black holes and the LIGOGW150914 and GW151226 detections of gravitational waves. In Section 6, we discuss the measurements of black hole masses, and we end with conclusions in Section 7. 

\section{Scalar-Tensor-Vector Gravity}

We adopt the metric signature $(+,-,-,-)$ and choose units with $c=1$. The MOG theory has a fully covariant action composed of scalar, vector and tensor fields~\cite{Moffat1}:
\begin{equation}
\label{action1}
S=S_G+S_\phi+S_S+S_M.
\end{equation}
The components of the action are the Einstein gravity action:
\begin{equation}
S_G=\frac{1}{16\pi}\int d^4x\sqrt{-g}\Big[\frac{1}{G}(R+2\Lambda)\Big],
\end{equation}
and the massive vector field $\phi_\mu$ action:
\begin{eqnarray}
S_\phi= - \int d^4x\sqrt{-g}\Big[\frac{1}{4}B^{\mu\nu}B_{\mu\nu}-V(\phi_\mu)\Big],
\end{eqnarray}
where $B_{\mu\nu}=\partial_\mu\phi_\nu-\partial_\nu\phi_\mu$ and $V(\phi_\mu)$ denotes a potential for $\phi_\mu$.  The action for the scalar fields $G$ and $\mu$ is
\begin{eqnarray}
\label{Saction}
S_S=\int d^4x\sqrt{-g}\Big[\frac{1}{G^3}\Big(\frac{1}{2}g^{\mu\nu}\nabla_\mu G\nabla_\nu G-V(G)\Big)
+\frac{1}{\mu^2 G}\Big(\frac{1}{2}g^{\mu\nu}\nabla_\mu\mu\nabla_\nu\mu - V(\mu)\Big)\Big].
\label{scalar}
\end{eqnarray}
Here, $\nabla_\mu$ denotes the covariant derivative with respect to the metric $g_{\mu\nu}$, $V(G)$ and $V(\mu)$ denote potentials for the fields
$G$ and $\mu$, respectively.  The energy-momentum tensor is defined as
\begin{equation}
T_{X\mu\nu}=-\frac{2}{\sqrt{-g}}\frac{\delta S_X}{\delta g^{\mu\nu}},\quad (X=[M,\phi,S]).
\end{equation}

The STVG field equations are given by~\cite{Moffat1}:
\begin{equation} 
\label{mog1}
G_{\mu\nu}-\Lambda g_{\mu\nu}+Q_{\mu\nu}=-8\pi GT_{\mu\nu},
\end{equation}
\begin{equation}
\label{mog2} 
\nabla_{\nu}B^{\mu\nu}+ \frac{\partial V(\phi)}{\partial \phi_\mu}=-J^\mu,
\end{equation}
\begin{equation}
\label{mog3}
\nabla_\sigma B_{\mu\nu}+\nabla_\nu B_{\sigma\mu}+\nabla_\sigma B_{\nu\mu}=0,
\end{equation}
\begin{equation}
\label{mog4}
\Box G=K(x),
\end{equation}
\begin{equation}
\label{mog5}
\Box\mu=L(x).
\end{equation}
We have
\begin{equation}
Q_{\mu\nu}=\frac{2}{G^2}(\nabla^\alpha G\nabla_\alpha Gg_{\mu\nu}-\nabla_\mu G\nabla_\nu G)-\frac{1}{G}(\Box Gg_{\mu\nu}-\nabla_\mu\nabla_\nu G).
\end{equation}
Moreover, 
\begin{equation}
K(x)=\biggl(\frac{16\pi}{3+16\pi}\biggr)\biggl[\frac{3}{8\pi G}(1+4\pi)\nabla^\alpha G\nabla_\alpha G 
- \frac{G}{2\mu^2}\Box\mu+\frac{1}{2}G^2\biggl(T+\frac{\Lambda}{4\pi G}\biggr)+\frac{1}{\sqrt{\alpha G_N}}T^{M\mu\nu}u_\nu\phi_\mu\biggr],
\end{equation} 
and
\begin{equation}
L(x)=\frac{1}{G}\nabla^{\alpha}G\nabla_{\alpha}\mu+\frac{2}{\mu}\nabla^\alpha\mu\nabla_\alpha\mu+\mu^2G\frac{\partial V(\phi_\mu)}{\partial \mu}.
\end{equation}
$G_{\mu\nu}$ is the Einstein tensor $G_{\mu\nu}=R_{\mu\nu}-\frac{1}{2}g_{\mu\nu}R$, $\Lambda$ is the cosmological constant, $\Box=\nabla^\mu\nabla_\mu$, $T=g^{\mu\nu}T_{\mu\nu}$ and $G$ and $\mu$ are scalar fields.  The Ricci curvature tensor is defined by
\begin{equation}
R_{\mu\nu}=\partial_\nu{\Gamma_{\mu\sigma}}^\sigma-\partial_\sigma{\Gamma_{\mu\nu}}^\sigma+\Gamma_{\mu\sigma}^\alpha\Gamma_{\alpha\nu}^\sigma
-\Gamma_{\mu\nu}^\alpha\Gamma_{\alpha\sigma}^\sigma.
\end{equation}
The potential $V(\phi_\mu)$ for the vector field $\phi_{\mu}$ is given by\footnote[2]{The scalar field $\omega(x)$ introduced in the original STVG paper is taken to be constant and $\omega=1$.}
\begin{equation}
\label{Vphi}
V(\phi_\mu)=-\frac{1}{2}\mu^2\phi^\mu\phi_\mu.
\end{equation}

The total energy-momentum tensor is defined by
\begin{eqnarray} 
T_{\mu\nu}=T^M_{\mu\nu}+T^\phi_{\mu\nu}+T^G_{\mu\nu}+T^\mu_{\mu\nu},
\end{eqnarray} 
where $T^M_{\mu\nu}$ is the energy-momentum tensor for the ordinary matter, and
\begin{equation}
\label{mog7}
T^\phi_{\mu\nu}=-\frac{1}{4\pi}\biggl[B_{\mu}^{~\alpha}B_{\nu\alpha}-g_{\mu\nu}\left(\frac{1}{4}B^{\rho\alpha}B_{\rho\alpha}+V(\phi_\mu)\right)
+2\frac{\partial V(\phi_\mu)}{\partial g^{\mu\nu}}\biggr],
\end{equation}
\begin{equation}
T^G_{\mu\nu}=-\frac{1}{4\pi G^3}\biggl(\nabla_\mu G\nabla_\nu G-\frac{1}{2}g_{\mu\nu}\nabla_\alpha G\nabla^\alpha G\biggr),
\end{equation}
\begin{equation}
T^\mu_{\mu\nu}=-\frac{1}{4\pi G\mu^2}\biggl(\nabla_\mu\mu\nabla_\nu\mu-\frac{1}{2}g_{\mu\nu}\nabla_\alpha\mu\nabla^\alpha\mu\biggr).
\end{equation}

The covariant current density $J^\mu$ for matter is defined by
\begin{equation}
\label{currentdensity}
J^\mu=\kappa T^{M\mu\nu}u_\nu,
\end{equation}
where $\kappa=\sqrt{\alpha G_N}$, $\alpha=(G-G_N)/G_N$ is a dimensionless scalar field, $G_N$ is Newton's constant, $u^\mu=dx^\mu/ds$ and $s$ is the proper time along a particle trajectory.  The perfect fluid energy-momentum tensor for matter is given by
\begin{equation}
\label{energymomentum}
T^{M\mu\nu}=(\rho_M+p_M)u^\mu u^\nu-p_Mg^{\mu\nu},
\end{equation}
where $\rho_M$ and $p_M$ are the density and pressure of matter, respectively, and for the fluid $u^\mu$ is the comoving four-velocity. We get from (\ref{currentdensity}) and (\ref{energymomentum}) by using $u^\nu u_\nu=1$:
\begin{equation}
J^\mu=\kappa\rho_M u^\mu.
\end{equation}
The gravitational source charge is given by
\begin{equation}
Q=\kappa\int d^3x J^0(x).
\end{equation}
From (\ref{mog2}) and (\ref{Vphi}) we get
\begin{equation}
\nabla_\mu J^\mu=\mu^2\nabla_\mu\phi^\mu.
\end{equation}
By requiring the condition $\nabla_\mu\phi^\mu=0$, we obtain 
\begin{equation}
\nabla_\mu J^\mu=0.
\end{equation}
The total density is given by
\begin{equation}
\rho=\rho_M+\rho_G+\rho_\phi+\rho_\mu.
\end{equation}

\section{Weak Field Approximation, Modified Acceleration Law and Gravitational Waves}

The weak field approximation is based on a perturbation about the Minkowski metric $\eta_{\mu\nu}$:
\begin{equation}
g_{\mu\nu}=\eta_{\mu\nu}+\lambda h_{\mu\nu},
\end{equation}
where we have set $\lambda=\sqrt{16\pi G}$ and the condition $g_{\mu\nu}g^{\mu\rho}={\delta_\nu}^\rho$ requires that $g^{\mu\nu}=\eta^{\mu\nu}-\lambda h^{\mu\nu}$.

The test particle equation of motion is given by
\begin{equation}
\label{eqmotion}
\frac{d^2x^\mu}{ds^2}+{\Gamma^\mu}_{\alpha\beta}\frac{dx^\alpha}{ds}\frac{dx^\beta}{ds}=\frac{q}{m}{B^\mu}_\nu\frac{dx^\nu}{ds},
\end{equation}
where $m$ and $q=\sqrt{\alpha G_N}m$ are the test particle mass and gravitational charge, respectively, and $\phi_\mu=(\phi_0,\phi_i)\,(i=1,2,3)$.  Note that $q/m=\sqrt{\alpha G_N}$, so that the inertial mass $m$ of the test particle cancels in the equation of motion (\ref{eqmotion}), so that the theory satisfies the (weak) equivalence principle for homogeneous gravitational fields.

Assuming that $V(\phi_\mu)$ is given by (\ref{Vphi}) and $\partial_\nu\phi^\nu=0$, the weak field spherically symmetric static, point particle solution for $\phi_0(r)$ is obtained from the equation ($\phi_0'=d\phi_0/dr$):
\begin{equation}
\phi_0''+\frac{2}{r}\phi_0'-\mu^2\phi_0=0,
\end{equation}
where $\mu$ is the mass of the vector field $\phi_\mu$. The solution is given by
\begin{equation}
\label{phisolution}
\phi_0(r)=-Q\frac{\exp(-\mu r)}{r},
\end{equation}
where the gravitational charge $Q=\sqrt{\alpha G_N}M$ and $M$ is the mass of the source particle.

In the slow motion and weak field approximation, $dr/ds\sim dr/dt$ and $2GM/r\ll 1$, and for the radial acceleration of the test particle we get~\cite{Moffat1}:
\begin{equation}
\frac{d^2r}{dt^2}+\frac{GM}{r^2}=\frac{qQ}{m}\frac{\exp(-\mu r)}{r^2}(1+\mu r).
\end{equation}
For $ qQ/m=\alpha G_NM$ and $G=G_N(1+\alpha)$, the modified Newtonian acceleration law for a point particle is given by
\begin{equation}
\label{accelerationlaw}
a(r)=-\frac{G_NM}{r^2}[1+\alpha-\alpha\exp(-\mu r)(1+\mu r)].
\end{equation} 
We can rewrite this modified acceleration equation as 
\begin{equation}
\label{effectiveaccel}
a(r)=-\frac{{\cal G}(r)M}{r^2},
\end{equation}
where the effective gravitational coupling strength is given by
\begin{equation}
\label{effectiveG}
{\cal G}(r)=G_N[1+\alpha-\alpha\exp(-\mu r)(1+\mu r)].
\end{equation}

For a continuous distribution of matter we get
\begin{equation}
\label{accelerationlaw2}
a({\vec x})=-G_N\int d^3x'\frac{\rho({\vec x}')({\vec x}-{\vec x}')}{|{\vec x}-{\vec x}'|^3}
[1+\alpha-\alpha\exp(-\mu|{\vec x}-{\vec x}'|)(1+\mu|{\vec x}-{\vec x}'|)].
\end{equation}
For a given density $\rho({\vec x})$ the potential is
\begin{equation}
\label{potential}
\Phi(\vec{x}) = - G_N \int d^3x'\frac{\rho(\vec{x}')}{|\vec{x}-\vec{x}'|}\Big[1+\alpha -\alpha \exp(-\mu|\vec{x}-\vec{x}'|)\Big].
\end{equation}
We can write the MOG acceleration law for a continuous distribution of matter as
\begin{equation}
a({\vec x})=-\int d^3x'\frac{{\cal G}({\vec x -\vec x'})\rho({\vec x}')({\vec x}-{\vec x}')}{|{\vec x}-{\vec x}'|^3},
\end{equation}
where 
\begin{equation}
\label{effectiveGdistribution}
{\cal G}({\vec x-\vec x'})=G_N[1+\alpha-\alpha\exp(-\mu|{\vec x}-{\vec x}'|)(1+\mu|{\vec x}-{\vec x}'|)].
\end{equation}

We define the field variable:
\begin{equation}
\gamma^{\mu\nu}=h^{\mu\nu}-\frac{1}{2}\eta^{\mu\nu}h,
\end{equation}
where $h=\eta^{\mu\nu}h_{\mu\nu}={h^\mu}_\mu$. The linearized metric field equations become:
\begin{equation}
\label{Fieldeqs}
\Box\gamma^{\mu\nu}=-\lambda T^{M\mu\nu},
\end{equation}
where $\Box=\partial^\alpha\partial_\alpha$, we have dropped the cosmological constant $\Lambda$, and we have adopted the condition:
\begin{equation}
\partial_\nu\gamma^{\mu\nu}=0.
\end{equation}
The linearized field equations for $G$ and $\mu$ are given by
\begin{equation}
\frac{1}{G}\Box G=\frac{8\pi G}{3+16\pi}T^M,
\end{equation}
and
\begin{equation}
\Box\mu=0.
\end{equation}

The retarded solution of (\ref{Fieldeqs}) is
\begin{equation}
\gamma^{\mu\nu}({\bf x},t)=-\frac{\lambda}{4\pi}\int d^3x'\frac{T^{M\mu\nu}(t-|{\bf x}-{\bf x}'|,{\bf x}')}{|{\bf x}-{\bf x}'|}.
\end{equation}
We have restricted the energy-momentum tensor to the matter tensor $T^{M\mu\nu}$ which satisfies the conservation law:
\begin{equation}
\partial_\nu T^{M\mu\nu}=0.
\end{equation}

In the radiation zone, far away from the matter system, we can replace $|{\bf x}-{\bf x}'|$ by $|{\bf x}|$ to give
\begin{equation}
\gamma^{\mu\nu}({\bf x},t)=-\frac{\lambda}{4\pi r}\int d^3x' T^{M\mu\nu}(t-r,{\bf x}'),
\end{equation}
where $r=|{\bf x}|$. For the calculation of the gravitational energy flux, we can in the radiation zone regard $\gamma^{\mu\nu}$ as a plane wave with only two transverse polarizations. Then, we have that
\begin{equation}
\gamma^{kl}({\bf x},t)=-\frac{\lambda}{24\pi r}{\ddot Q}^{kl},
\end{equation}
where the dots stand for time derivatives, $k,l=1,2,3$ and the right-hand side is to be evaluated at the retarded time $t-r$. The total power of tensor gravitational wave energy in the generalized theory is
\begin{equation}
\label{Power}
P=\frac{{\cal G}}{45 c^5}{\dddot Q}_{kl}{\dddot Q}_{kl},
\end{equation}
where $Q_{kl}$ is the quadrupole moment:
\begin{equation}
Q_{kl}=\int d^3x'\biggl(3x'_kx'_l-r^{'2}\delta_{kl}\biggr)\rho_M({\vec  x}').
\end{equation}
The gravitational waves are a result of the accelerated motion of masses, so for weak gravitational fields we use the modified effective acceleration law (\ref{effectiveaccel}). We have replaced in (\ref{Power}) the gravitational coupling strength $G$ by the effective gravitational coupling ${\cal G}$ in (\ref{effectiveG}), in order to account for the repulsive effect of the vector field $\phi_\mu$ on the accelerated motion of a massive source. An analysis of the field equations (\ref{mog2}) and (\ref{mog3}) for weak gravitational fields reveals that the vector field $\phi_\mu$ does not produce dipole radiation, because $Q=\sqrt{\alpha G_N}M > 0$ and due to the conservation of the gravitational source charge, ${\dot Q}=0$ ($\dot M=0$), there is no monopole radiation.

For two particles or two spherical masses moving in elliptical orbits about their common center-of-mass, the time-averaged power radiated by the system is ~\cite{PetersMatthews,Peters}:
\begin{equation}
\langle P\rangle\equiv \biggl\langle\frac{dE}{dt}\biggr\rangle = \frac{32}{5}\frac{{\cal G}^4}{5c^5}\frac{(m_1m_2)^2(m_1+m_2)}{a^5(1-e^2)^{7/2}}\biggl(1+\frac{73}{24}e^2+\frac{37}{96}e^4\biggr),
\end{equation}
where $m_1,m_2$ are the masses, $a$ is the semi-major axis, $e$ is the eccentricity. We have used the generalized Kepler's third law for the orbital angular velocity:
\begin{equation}
\omega=\frac{[{\cal G}(R)(m_1+m_2)a(1-e^2)]^{1/2}}{R^2},
\end{equation}
where $R$ is the distance between the binary components.

We have ${\cal G}(R)\rightarrow G_N$ for $r=R\ll\mu^{-1}$, where $\mu^{-1}\sim 24$ kpc from the fits to galaxy rotation curves and cluster dynamics~\cite{MoffatRahvar1,MoffatRahvar2}. Thus, for well-separated orbiting compact bodies, the time-averaged tensor gravitational wave emission power is given by
\begin{equation}
\label{WeakGravPower}
\langle P\rangle\equiv \biggl\langle\frac{dE}{dt}\biggr\rangle = \frac{32}{5}\frac{G_N^4}{5c^5}\frac{(m_1m_2)^2(m_1+m_2)}{a^5(1-e^2)^{7/2}}\biggl(1+\frac{73}{24}e^2+\frac{37}{96}e^4\biggr).
\end{equation}

As the binary system loses energy by gravitational radiation, the orbital period $P_b$ of a binary system of compact objects decreases as
\begin{equation}
{\dot P}_b=-\frac{192\pi G_N^{5/3}}{5c^5}\biggl(\frac{P_b}{2\pi}\biggr)^{-5/3}(1-e^2)^{-7/2}\biggl(1+\frac{73}{24}e^2+\frac{37}{96}e^4\biggr)m_1m_2(m_1+m_2)^{-1/3}.
\end{equation}
The PPN formalism determined by expansions in $v/c$ yields corrections to the gravitational radiation formulas~\cite{Blanchet}. The binary pulsar PSR B1913+16 data agree with the GR prediction of ${\dot P}_b$ to within $\sim 0.2\%$. For the most relativistic binary systems the observed rate of change of the period agrees with GR to better than $0.03\%$.

\section{Generalized Kerr Black Hole Solution}

An exact generalized Kerr solution of the STVG field equations has been derived~\cite{Moffat5,Moffat6,Moffat7}. The field equations for the special case $G\sim {\rm constant}$ and $Q=\sqrt{\alpha G_N}M\sim {\rm constant}$, ignoring in the present universe the small $\phi_\mu$ field mass $m_\phi\sim 10^{-28}$ eV, are given by
\begin{equation}
\label{phiFieldEq}
R_{\mu\nu}=-8\pi GT^\phi_{\mu\nu},
\end{equation}
\begin{equation}
\label{Bequation}
\nabla_\nu B^{\mu\nu}=\frac{1}{\sqrt{-g}}\partial_\nu(\sqrt{-g}B^{\mu\nu})=0,
\end{equation}
\begin{equation}
\label{Bcurleq}
\nabla_\sigma B_{\mu\nu}+\nabla_\mu B_{\nu\sigma}+\nabla_\nu B_{\sigma\mu}=0.
\end{equation}
The energy-momentum tensor ${{T^\phi}_\mu}^\nu$ is 
\begin{equation}
\label{Tphi}
{{T^\phi}_\mu}^\nu=-\frac{1}{4\pi}({B_{\mu\alpha}}B^{\nu\alpha}-\frac{1}{4}{\delta_\mu}^\nu B^{\alpha\beta}B_{\alpha\beta}).
\end{equation}

The generalized Kerr black hole solution metric is given by
\begin{equation}
\label{KerrMOG}
ds^2=\frac{\Delta}{\rho^2}(dt-a\sin^2\theta d\phi)^2-\frac{\sin^2\theta}{\rho^2}[(r^2+a^2)d\phi-adt]^2-\frac{\rho^2}{\Delta}dr^2-\rho^2d\theta^2,
\end{equation}
where
\begin{equation}
\Delta=r^2-2GMr+a^2+\alpha(1+\alpha) G_N^2M^2,\quad \rho^2=r^2+a^2\cos^2\theta.
\end{equation}
The spacetime geometry is axially symmetric around the $z$ axis. Horizons are determined by the roots of $\Delta=0$:
\begin{equation}
r_\pm=G_N(1+\alpha)M\biggl[1\pm\sqrt{1-\frac{a^2}{G_N^2(1+\alpha)^2M^2}-\frac{\alpha}{1+\alpha}}\biggr].
\end{equation}
An ergosphere horizon is determined by $g_{00}=0$:
\begin{equation}
r_E=G_N(1+\alpha)M\biggl[1+\sqrt{1-\frac{a^2\cos^2\theta}{G_N^2(1+\alpha)^2M^2}-\frac{\alpha}{1+\alpha}}\biggr].
\end{equation}
The solution is fully determined by the Arnowitt-Deser-Misner (ADM) mass $M$ and spin parameter $a$ ($a=cS/GM^2$ where $S$ denotes the spin-angular momentum) measured by an asymptotically distant observer. When $a=0$ the solution reduces to the generalized Schwarzschild black hole metric solution:
\begin{equation}
\label{MOGmetric}
ds^2=\biggl(1-\frac{2G_N(1+\alpha)M}{r}+\frac{\alpha(1+\alpha) G_N^2M^2}{r^2}\biggr)dt^2-\biggl(1-\frac{2G_N(1+\alpha)M}{r}+\frac{\alpha(1+\alpha) G_N^2M^2}{r^2}\biggr)^{-1}dr^2-r^2d\Omega^2.
\end{equation}
When the parameter $\alpha=0$ the generalized solutions reduce to the GR Kerr and Schwarzschild black hole solutions. Both the generalized Kerr black hole and static spherically symmetric black hole solutions are algebraically equivalent to the Kerr-Newman and Reissner-Norstr\"om black hole solutions~\cite{Moffat5,Moffat6,Moffat7}.

\section{Black Hole Binary and LIGO Gravitational Wave Detection}

The recent detection of a gravitational wave from the inspiral of binary black hole systems~\cite{Abbott1,Abbott2} opens a new era in observational astronomy. The existence of gravitational waves was inferred from observations of binary pulsar systems~\cite{TaylorWeisberg,HulseTaylor}. For the first time the direct detection of gravitational waves has made it possible to observe a black hole-black hole (BH-BH) merger and infer its parameters independently of electromagnetic observations. This provides an unprecedented opportunity to study two-body motion of a compact-object binary in the large velocity, strong gravitational nonlinear regime. We can witness the final merger of the BH-BH system and determine whether the GW150914 and GW151226 events are consistent with the binary black hole solution in GR. A study of this problem has been carried out~\cite{LIGOVirgo1} and it was claimed that the GW150914 data is consistent with GR. They found that the final remnant black hole mass and spin, determined from the inspiral and coalescent phases of the detected signal are compatible with the GR solution. 

The GW150914 source lies at a luminosity distance of $410^{+160}_{-180}$ Mpc corresponding to a redshift $z=0.09^{+0.03}_{-0.04}$. The inferred initial black hole masses are  $m_1=36^{+5}_{-4} M_\odot$ and $m_2=29^{+4}_{-4}M_\odot$, the final black hole mass is $M=62^{+4}_{-4}M_\odot$, with $3.0^{+0.5}_{-0.5}M_\odot c^2$ energy radiated away in gravitational waves, and the final black hole spin inferred from GR is $a=0.67^{+0.05}_{-0.07}$. The gravitational wave luminosity determined from GR reached a peak value of $3.6^{+0.5}_{-0.4}\times 10^{56}$ erg/s equivalent to $200^{+30}_{-20}M_\odot c^2/s$. The GW151226 source lies at a luminosity distance  $440^{+180}_{-190}\,{\rm Mpc}$ corresponding to a redshift $z=0.09^{+0.03}_{-0.04}$. The inferred masses are $m_1=14.2^{+8.3}_{-3.7}$ and $m_2=7.5^{+2.3}_{-2.3}$ and the final black hole mass is $M=20.8^{+6.1}_{-1.7}$. The gravitational wave detection events point to them being produced by the coalescence of two black holes - their orbital inspiral and merger and final black hole ringdown. During the period of 0.2 s, the GW150914 detected signal increases in frequency and amplitude from about 8 cycles from 35 Hz to a maximum 150 Hz. The merging of the black holes requires a numerical solution of the GR field equations. This has been accomplished for GR~\cite{Pretorius1,Pretorius2,Pretorius3} and solutions have been derived that can match the gravitational wave form signals. Future work will require that numerical solutions to the generalized gravitational field equations be obtained, leading to the accurate determination of gravitational wave forms. 

In our gravity theory and in GR, there are two independent gravitational wave polarization strains, $h_+(t)$ and $h_\times(t)$. The polarization strains during the inspiral of the black holes can be written as~\cite{LIGOVirgo2}:
\begin{equation}
h_+(t)=A_{GW}(t)(1+\cos^2\iota)\cos(\phi_{GW}(t)),
\end{equation}
\begin{equation}
h_\times(t)=-2A_{GW}(t)\cos\iota\sin(\phi_{GW}(t)),
\end{equation}
where $A_{GW}(t)$ and $\phi_{GW}(t)$ denote the amplitude and phase, respectively, and $\iota$ is the inclination angle. Post-Newtonian theory is used to compute $\phi_{GW}(t,m_{1,2},S_{1,2})$ where $S_1$ and $S_2$ denote the black hole spins, and the perturbative expansion is in powers of $v/c\sim 0.2-0.5$. A description of the gravitational wave phase is
\begin{equation}
\phi_{\rm GW}(t)\sim 2\pi\biggl(ft+\frac{1}{2}{\dot f}t^2\biggr)+\phi_0,
\end{equation}
where $f$ is the gravitational wave frequency. The strain $h_+(t)$ can be expressed in the generalized theory as a rough estimate:
\begin{equation}
\label{straineq}
h_+(t)\sim\frac{{\cal G}^2(R)m_1m_2}{DR(t)c^4}(1+\cos^2\,\iota)\cos\biggl(\int^t f(t')dt'\biggr),
\end{equation}
where $D$ is the distance to the binary system source, ${\cal G}(R)$ is the effective gravitational strength (\ref{effectiveG}) and $R(t)$ is the radial distance of closest approach during the inspiraling merger.  We have
\begin{equation}
\label{strainfeq}
f(t)=\frac{5^{3/8}}{8\pi}\biggl(\frac{c^3}{{\cal G}{\cal M}_c}\biggr)^{5/8}(t_{\rm coal}-t)^{-3/8},
\end{equation}
where ${\cal M}_c$ is the chirp mass:
\begin{equation}
\label{chirpmass}
{\cal M}_c=\frac{(m_1m_2)^{3/5}}{(m_1+m_2)^{1/5}}=\frac{c^3}{{\cal G}}\biggl[\frac{5}{96}\pi^{-8/3}f^{-11/3}{\dot f}\biggr]^{3/5}.
\end{equation}
The characteristic evolution time at the frequency $f$ is
\begin{equation}
\label{tevolveeq}
t_{\rm evol}\equiv\frac{f}{{\dot f}}=\frac{8}{3}(t_{\rm coal}-t)=\frac{5}{96\pi^{8/3}}\frac{c^5}{f^{8/3}({\cal G}{\cal M}_c)^{5/3}},
\end{equation}
and the chirp ${\dot f}$ is given by~\cite{Buonanno,Duez}:
\begin{equation}
\label{chirp}
{\dot f}=\frac{96}{5}\frac{c^3f}{{\cal G}{\cal M}_c}\biggl(\frac{\pi f}{c^3}{\cal G}{\cal M}_c\biggr)^{8/3}.
\end{equation}

Observed black holes come in two classes: the supermassive black holes at the centers of galaxies and stellar-size black holes. The former are in the mass range $10^5M_\odot-10^{10}M_\odot$, while the latter are in the mass range $2.5-10M_\odot$. The fact that the black hole masses inferred from the GW150914 signal strength have the values $m_1\sim 36M_\odot$ and $m_2\sim 29M_\odot$ leads to the problem of how such massive, intermediate black hole binaries could be formed. The mass wind and metallicity $Z$ of progenitor stars generally conspire to lead through collapse to black holes with the mass $\lesssim 10M_\odot$. 

In our generalized gravitation theory, we can produce a solution to this black hole evolution problem. For initial well-separated binary black holes in the weak gravitational field regime, ${\cal G}(R)\sim G_N$ according to the effective gravitational field strength (\ref{effectiveG}) for $R << \mu^{-1}\sim 24$ kpc (determined by fits to galaxy rotation curves and cluster dynamics~\cite{MoffatRahvar1,MoffatRahvar2}). As the two black holes, described by the Kerr-MOG and Schwarzshild-MOG solutions, coalesce and merge to the final black hole with $G\sim G_N(1+\alpha)$, we can choose the parameter $\alpha$ in a range of values of order unity. 

As the two black holes merge the masses, the gravitational charges and the spins also merge to their final values for the quiescent black hole after the ringdown phase. During this stage the repulsive force exerted on the two black holes, due to the vector field charges $Q_1=\kappa m_1$ and $Q_2=\kappa m_2$, decreases to zero and $G=G_N(1+\alpha)$ and $Q=\kappa M$ where $\alpha$ and $M$ are the final values of the quiescent black hole mass and $\alpha$. During the rapid coalescing strong gravity phase, the repulsive vector force only partially cancels the attractive force and $\alpha$ and $G\sim G_N(1+\alpha)$. In this merging phase, we have ${\cal G}\sim G_N(1+\alpha)$ and we obtain
\begin{equation}
\label{straineq2}
h_+(t)\sim\frac{G^2_N(1+\alpha)^2m_1m_2}{DR(t)c^4}(1+\cos^2\,\iota)\cos\biggl(\int^t f(t')dt'\biggr),
\end{equation}
\begin{equation}
\label{chirpmass2}
{\cal M}_c=\frac{(m_1m_2)^{3/5}}{(m_1+m_2)^{1/5}}=\frac{c^3}{G_N(1+\alpha)}\biggl[\frac{5}{96}\pi^{-8/3}f^{-11/3}{\dot f}\biggr]^{3/5},
\end{equation}
\begin{equation}
\label{chirp2}
{\dot f}=\frac{96}{5}\frac{c^3f}{G_N(1+\alpha){\cal M}_c}\biggl(\frac{\pi f}{c^3}G_N(1+\alpha){\cal M}_c\biggr)^{8/3}.
\end{equation}

The increase of $G$ in the final stage of the merging of the black holes can lead to a fitting of the LIGO data for binary black holes in agreement with the observed values of stellar-mass binary black holes. We have for the GR chirp mass using $m_1=36M_\odot$ and $m_2=29M_\odot$ the value ${\cal M}_{c\rm GR}=28.0M_\odot$, and from $G_N(1+\alpha){\cal M}_{c\rm MOG}=G_N{\cal M}_{c\rm GR}$ we get
\begin{equation}
\alpha=\frac{{\cal M}_{c\rm GR} - {\cal M}_{c\rm MOG}}{{\cal M}_{c\rm MOG}}.
\end{equation}

Estimating $f$, $h_+$ and ${\dot f}$ from the LIGO data, we obtain in Table 1 the values for $\alpha$ for different BH-BH component masses.  In the detector frame the total mass is $M_{\rm GR}$ and $M_{\rm MOG}$, less the mass lost by gravitational radiation. The bounds on the sum of the Schwarzschild radii of the black hole binary components are $r_{\rm sGR}=2G_NM_{\rm GR}$ and $r_{\rm sMOG}=2G_N(1+\alpha)M_{\rm MOG}$. To reach an orbital frequency 75 Hz that is half the gravitational wave frequency the orbiting black holes were very close Newtonian point masses only $\lesssim 350$ km apart. 

\begin{table}
\caption{Summary of values of $\alpha$, $m_1,m_2$ and chirp mass ${\cal M}_c$ for BH-BH binary systems and the GW LIGO events GW 150914 and GW 152612.}
\begin{center}
    \begin{tabular}{| l | l | l | l | l |}
    \hline
    Merging systems & $\alpha$ & $m_1(M_\odot)$ & $m_2(M_\odot)$ & ${\cal M}_c(M_\odot)$\\
    \hline
    GR GW150914 & $0$ & $36$ & $29$ & $28$\\
    GR GW151226 & $0$ & $14$ & $8$ & $8.9$\\
   	MOG GW150914 & $2.6$ & $10$ & $8$ & $7.8$\\ 
     MOG GW151226 & $2.0$ & $4.7$ & $3$ & $3$\\
   	MOG GW150914 & $5.7$ & $6$ & $4$ & $4.2$\\
     MOG GW150914 & $8.3$ & $4$ & $3$ & $3$\\ \hline
    \end{tabular}
\end{center}
\label{datasum}
\end{table}

The initial orbiting black holes are well-separated in distance from one another, so the weak gravitational wave power emission is given by (\ref{WeakGravPower}) with the appropriate PPN corrections included. The GR and MOG masses $m_1$ and $m_2$ of the two black holes are given in Table 1, and in this initial phase of the slowly inspiralling black holes no gravitational wave signal will be detected.  In the final merging stage of the black holes the value of ${\cal G}\sim G_N$, obtained from  the weak gravitational and slow velocity formula (\ref{effectiveG}) is no longer valid for strong gravitational fields. Thus, with $\alpha > 0$, we can fit the audible chirp signal LIGO data with smaller values for $m_1$ and $m_2$. As the black holes coalesce, the final quiescent MOG black hole will have a total mass $M=m_1+m_2$, less the amount of mass-energy emitted by gravitational wave emission, and it will be described by the Kerr-MOG metric (\ref{KerrMOG}). 

The GW150914 observed spin is determined by the effective spin parameter:
\begin{equation}
\chi_{\rm eff}=\frac{c}{GM}\biggl(\frac{{\vec S}_1}{m_1}+\frac{{\vec S}_2}{m_2}\biggr)\cdot {\hat{\vec L}}=\frac{m_1{\vec a}_1+m_2{\vec a}_2}{m_1+m_2}
\cdot{\hat{\vec L}},
\end{equation}
where ${\vec a}_1$ and ${\vec a}_2$ are the dimensionless vector component spin parameters and ${\hat{\vec L}}$ is the direction of orbital angular momentum. The observed value is $\chi_{\rm eff}=-0.06^{+0.17\pm 0.01}_{-0.18\pm 0.07}$~\cite{LIGOVirgo2}. The low value of the spin parameter $\chi_{\rm eff}$ has been found to be in tension with stellar astrophysics and progenitor models of black hole formation~\cite{Zaldarriaga}. For the MOG black hole ${\vec a}=c{\vec S}/G_N(1+\alpha)M^2$, so for $\alpha > 0$ the decrease in the magnitude of ${\vec a}$ compared to the GR prediction can bring the value of $\chi_{\rm eff}$ closer to its low observed value for models of binary black hole formation.

\section{Conclusions}

Black holes play an important role in astrophysical phenomena, ranging from binaries to ultra-luminous X-ray sources (ULXs), galaxies and quasars. Stellar-mass black holes offer us the best opportunity to investigate these objects in detail. The relative nearness and time-scale variability allow us to study their electromagnetic properties through their accretion regimes. Accurate determinations of black hole masses is critical to test models of massive progenitors. Although solitary black holes do not emit electromagnetic radiation, they become detectable X-ray sources when they have a stellar companion transferring matter to them. The robust method of measuring stellar masses uses Kepler's third law of motion. This determines the orbital period and the radial velocity semi-amplitude of the companion star. These two quantities combine to determine the mass function equation~\cite{Casares,Remillard}:
\begin{equation}
f(M)=\frac{K_c^3P_{\rm orb}}{2\pi G_N}=\frac{M_x^3\sin^3\iota}{(M_x+M_c)^2}=\frac{M_x\sin^3\iota}{(1+q_b)^2},
\end{equation}
where $P_{\rm orb}$ denotes the orbital period, $M_x$ the black hole mass, $M_c$ the mass of the companion star, $K_c$ the radial velocity semi-amplitude of the companion star, $\iota$ the inclination angle and $q_b=M_c/M_x$ the binary mass ratio. Observed stellar black hole masses range from $3M_\odot$ to $15M_\odot$. The formation of the black holes follows an evolutionary process depending on the initial mass of the progenitor, how much mass is lost during its evolution and on the supernova explosion mechanism. Mass is lost through stellar winds and the amount lost depends on the metallicity of the progenitor star. For a star of low metallicity $\sim 0.01$ of the solar metallicity, it is possible to end up with a black hole of $\lesssim 100M_\odot$, although there is no direct observational evidence available at present to support this. It is also conjectured that more massive black holes can be produced in the dense stellar populations of globular clusters. 

The system M33 X-7 harbors the heaviest star ever discovered in an X-ray binary orbiting the most massive black hole found in one of these systems. A $70M_\odot$ star orbits a $\sim 15M_\odot$ black hole every 3.45 days~\cite{Orosoz,Valsecchi}. The tight orbit and the massive components make M33 X-7 an evolutionary challenge. Finding a plausible evolutionary model has been complicated by the low luminosity of the stellar component - lower than is predicted for a single star of mass $70M_\odot$. 

Until observations confirm the existence of stellar black holes with a mass $36M_\odot$ and $18M_\odot$~\cite{Abbott3,Belczynski}, we can consider that the generalized gravitation theory fitting of the GW150914 data with black hole masses $\lesssim 10M_\odot$ to be a challenge for the future testing of the strong gravity regime of black holes. In particular, further research on the evolution of black hole binaries is needed to determine whether the evolution can support a binary stellar mass black hole with $M\sim 36M_\odot$. It has been conjectured that primordeal black holes could be produced in the early universe with mass $ > 10M_\odot$. However, such primordeal black holes have not been observed.

The role of gravitational waves in a modified gravitational theory has been investigated, and the radiated gravitational wave power has been derived. It is argued from the motion of compact bodies in the weak field $(2GM/c^2r\ll 1)$ and slow motion approximation that the repulsive force due to the vector field potential cancels the enhanced gravitational force between compact bodies, yielding the Newtonian acceleration law $G_NM/r^2$ together with PPN corrections and agreement with binary pulsar observational data. The photon propagates through a conformal metric along a null geodesic path and is screened for dense bodies such as the sun, yielding agreement with the solar system bending of light and Cassini probe experiments~\cite{Moffat2}. The speed of electromagnetic waves (massless photons) and gravitational waves (massless gravitons) is the speed of light. For weak gravitational fields the tensor radiated power reduces to the GR result for well-separated binary particles in agreement with the experimental results for binary pulsars. 

The generalized gravitational theory is applied to the gravitational wave detections by LIGO~\cite{Abbott1,Abbott2}. The field equations are restricted to the special case when $G=G_N(1+\alpha)\sim {\rm constant}$ and the small vector particle mass $m_\phi$ is neglected. We choose for the initial inspiralling phase of the black hole merger with well-separated point mass particles, two black hole component masses $m_1,m_2\leq 10M_\odot$, and the effective coupling strength ${\cal G}\sim G_N$.  As the two black holes merge, the weak field approximation is no longer valid and for the strong gravitational field, coalescing phase $G\sim G_N(1+\alpha)$. To fit the LIGO signal data, we can choose a range of values of the parameter $\alpha$. For the modified gravity $G_N(1+\alpha){\cal M}_{c\rm MOG}={G_N\cal M}_{c\rm GR}$, and values of the chirp mass and the frequency $f$ and its time derivative ${\dot f}$ are obtained from fits to the merging black hole LIGO signal data. This allows for a fit to the LIGO wave form signal data. After the ringdown phase, when the merger end-product relaxes to the final stationary, equilibrium MOG black hole, it will be described by the Kerr-MOG black hole solution that only depends on its final mass and spin.

It is argued that measurements of black hole masses by observations of X-ray binaries give masses $\lesssim 10M_\odot$, and not the masses $m_1=36M_\odot$, $m_2=29M_\odot$ and $m_1=14M_{\odot}$ and $m_2=7.5M_{\odot}$ (the mass $m_2$ is consistent with observed binary stellar black hole masses) inferred from the LIGO signals assuming the validity of GR. The observed upper bound $M\sim 10M_\odot$ on binary black holes masses is in accord with the evolutionary formation models of black holes and collapse models based on supernova explosions. Thus, the generalized gravity results we have obtained for the black hole masses are more in agreement with the observational data and theoretical model calculations currently available for the masses of stellar mass black holes than the inferred masses derived from GR and the LIGO data. A full numerical calculation of the STVG field equations is required to obtain an accurate derivation of the final strong field merging of the MOG black holes and the wave forms that fit the LIGO-Virgo signal data.

Another important prediction of our modified gravitational theory is the size of the black hole shadow predicted by MOG compared to that predicted by GR~\cite{Moffat6,Broderick}. The event horizon telescope (EHT) will be able to determine the size of the shadow cast by photons trapped by the strong gravitational field of the supermassive black hole Sagittarius A* (Sgr A*) with an error of about $5\%$, or an angular radius error of $1.5\,\mu {\rm as}$. The predicted size of the shadow for the Schwarzschild-MOG and Kerr-MOG black holes can determine a bound on $\alpha$ for Sgr A*.  For a Schwarzschild-MOG black hole an approximate formula for the size of the shadow is
\begin{equation}
r_{\rm shad}\sim (5.18+4\alpha)r_g,
\end{equation}
where $r_g=G_NM/c^2$ and $M=4.23\times 10^6M_\odot$. The angular radius $R=r_{\rm shad}/D$ where $D=8.3$ kpc is $R\sim 26\,\mu{\rm as}$ for $\alpha=0$ and $R\sim 46\,\mu{\rm as}$ for $\alpha=1$. This generalized gravity prediction of the size of the Sgr A* black hole shadow in conjunction with bounds on $\alpha$ obtained from future gravitational wave experimental results can distinguish for strong gravitational fields our generalized gravitational theory from general relativity.

\section*{Acknowledgments}

I thank Luis Lehner, Avery Broderick, Martin Green and Viktor Toth for helpful discussions. Research at the Perimeter Institute for Theoretical Physics is supported by the Government of Canada through industry Canada and by the Province of Ontario through the Ministry of Research and Innovation (MRI).

\end{document}